\documentclass[11pt,twoside]{article}

\usepackage{asp2006}
\usepackage{epsf}
\usepackage{psfig}
\usepackage{graphicx}

\begin{document}
\title{The Co-evolution of Galaxies, Black Holes, and AGN in a Hierarchical Universe}   
\author{Rachel S. Somerville}   
\affil{Max-Planck Institute for Astronomy}    

\begin{abstract} 
The observational link between Supermassive Black Holes (SMBH) and
galaxies at low redshift seems to be very tight, and statistically the
global evolution of star formation activity and BH accretion activity
also seem to trace each other closely. However, pinning down the
co-evolution of galaxies and BH on an object-by-object basis remains
elusive. I present results from new models for the joint evolution of
galaxies, SMBH, and AGN, which may be able to help resolve some of the
observational puzzles. A unique aspect of these models is our
treatment of self-regulated BH growth based on hydrodynamic
simulations of galaxy-galaxy mergers. Although these models do quite
well at reproducing the observed evolution of galaxies, they do not
reproduce the observed history of BH accretion, predicting too much
early accretion and not enough at late times. I suggest two possible
resolutions to this problem.

\end{abstract}



\section{Introduction}

There is strong evidence that the present-day properties of
Supermassive Black Holes (SMBH) and their host galaxies are tightly
linked --- for example we observe a tight relationship between BH mass
and the mass of the host spheroid in nearby galaxies
\citep[e.g.][]{haering:04}. However, we do not yet know whether this
BH-galaxy mass relationship evolved over cosmic time, or how galaxies
and SMBH arrived onto it. Perhaps most importantly, we do not yet
understand the physical origin of this relationship.

Mergers have often been proposed as a mechanism that can drive gas
into the nuclei of galaxies and fuel both starbursts and black hole
accretion. \citet{hopkins:05} used a large suite of hydrodynamic
simulations of galaxy mergers to characterize the episode of black
hole growth and the AGN accretion ``lightcurves'' in these
events. \citet{hopkins:06} then used this model to test whether
observed galaxy merger rates, the build-up of red/spheroidal galaxies,
and AGN luminosity functions are consistent with the picture in which
mergers both trigger AGN activity and cause morphological and color
transformation. They concluded that there is excellent statistical
agreement between all of these quantities. 

However, establishing a direct link between mergers and AGN activity
has proven controversial. Low luminosity AGN, which are more common
and can be studied in more detail at relatively low redshift ($z<1$),
do not show strong evidence for being predominantly in morphologically
disturbed hosts or close pairs \citep[e.g.][]{li:06,pierce:06}. Higher
luminosity quasars seem to be more often (though not always)
associated with disturbed hosts or to have close companions
\citep{bahcall:97,bennert:08,letawe:08}, but the samples with high
resolution imaging are small. It may be the case that AGN are most
easily identified in the very late stages of mergers, when the two
nuclei have completely coalesced and most signs of morphological
disturbance have become invisible. Especially at high redshift, at the
typically available image sensitivity, these very late stage merger
remnants might well appear to be ``normal'' spheroidal-type galaxies.

Of course it is also possible, even likely, that there are multiple
modes of BH growth. There are ample suggestions in the literature that
AGN activity could also be fed by bars or by stochastic accretion of
cold gas in galactic nuclei. \citet{hopkins_hernquist:06} presented a
model for fueling of BH in isolated disks by stochastic accretion, and
argued that while this fueling mode may be important in lower
luminosity objects and at late times, most of the growth of today's
massive BH occurred in the merger-driven mode at high redshift.

Another sign of co-evolution is that the global histories of star
formation and BH accretion seem to trace each other remarkably closely
over the whole range of lookback times over which these quantities
have robust observational estimates. The global star formation rate
density is almost exactly a factor of 2000 times the BH accretion rate
density, from redshift $z\sim5$ to the present. Moreover, even when
divided by mass or accretion rate, ``matched'' populations of galaxies
and black holes at $z<1$ also trace one another's activity (Zheng et
al. in prep). However, we know that at least at $z<1$, the bulk of the
star formation activity is associated with \emph{isolated} disks, not
mergers or spheroids \citep{bell:05}. Therefore, it is also strange
that these two kinds of activity trace each other, but do not seem to
be occuring in the same types of objects (Zheng et al. in prep).

At higher redshift, there may be a larger mismatch between between the
two kinds of activity. \citet{faucher-giguere:08} obtained constraints
on the global star formation density at $2<z<4.2$ from the
Lyman-$\alpha$ forest opacity in QSO spectra. They found that the
hydrogen photoionization rate, and hence the star formation rate
density, was remarkably flat over this redshift range, in contrast to
the sharply peaked QSO luminosity density, which falls off sharply at
$z>2$. This may indicate that there is a time delay between SF
activity and QSO activity, which is naturally explained in the merger
picture (any galaxy with cold gas can form stars, but galaxies have to
``wait'' for a major merger before significant accretion onto the BH
can occur).

One way to test this picture, in which mergers are responsible for
triggering both AGN activity and morphological and spectrophotometric
transformation of galaxies, is to build cosmological models that treat
the growth of galaxies, black holes, and AGN self-consistently. In
addition, many astronomers now believe that the energy released by
accreting black holes may play a crucial role in regulating galaxy
formation.  Here we describe one such semi-analytic model for the
joint formation of galaxies, black holes, and AGN, and present some
predictions from these models.

\section{Unified Models for Galaxies, Black Holes, and AGN}

The foundations of our model, which pertain to the growth of structure
in the dark matter component, and the formation of galaxies, are
described in \citet{sp:99}, \citet{spf:01}, and subsequent
works. Briefly, the models are implemented within dark matter halo
``merger trees'', and include approximate treatments for atomic
cooling of gas, photoionization, the formation of angular momentum
supported disks (including estimates of their sizes and circular
velocities), star formation according to an empirical ``Kennicutt''
Law, supernova feedback, and chemical evolution. We have recently
implemented new machinery within this framework to treat black hole
growth and the associated AGN feedback. Our new models are fully
described in \citet[][S08]{somerville:08}. Here we give a very brief
synopsis of the most important new ingredients. The models presented
here are for a ``Concordance'' $\Lambda$CDM cosmology with
$\Omega_m=0.3$, $\Omega_{\Lambda}=0.7$, $H_0=70$ km/s/Mpc,
$\sigma_8=0.9$, $n_s=1$.

\subsubsection{Galaxy-galaxy Mergers}

We assume that galaxy mergers may induce a burst of star formation and
destroy any pre-existing disk component, depending on the mass ratio
of the smaller to the larger galaxy, $\mu$. We parameterize the
strength and timescale of these bursts according to the results of a
large suite of hydrodynamic simulations of galaxy mergers
\citep{robertson:06,cox:08}, as described in S08. Mergers can also
heat and thicken, or even destroy, a pre-existing disk component,
driving galaxies towards morphologically earlier Hubble types.  We
assume that the fraction of the pre-existing stars that is transferred
from a disk to a spheroidal component is a strongly increasing
function of the merger mass ratio $\mu$, such that minor mergers with
$\mu < 0.2$ have little effect, and major mergers with $\mu > 0.25$
leave behind a spheroid-dominated remnant.

\subsubsection{Bright Mode AGN and AGN-driven winds}

In our model, mergers also trigger the accretion of gas onto
supermassive black holes in galactic nuclei. Each top-level halo in
our merger trees is seeded with a black hole of mass $M_{\rm seed}
\simeq 100 M_{\odot}$. Following a merger, the black hole is allowed
to grow at the Eddington rate until the BH reaches a critical mass,
where the radiative energy being emitted by the AGN becomes sufficient
to halt further accretion. This self-regulated treatment of black hole
growth is based on hydrodynamic simulations including BH growth and
feedback \citep{springel:05a,dimatteo:05,hopkins_bhfpth:07}, and is
described in more detail in S08.  Energy radiated by black holes
during this ``bright'', quasar-like mode can also drive galactic-scale
winds, clearing cold gas from the post-merger remnants
\citep{springel:05b}. Our model for momentum-driven AGN winds is
described in S08.

\subsubsection{Radio Mode Feedback}

In addition to the rapid growth of BH in the merger-fueled,
radiatively efficient ``bright mode'' described above, we assume that
BH also experience a low-Eddington-ratio, radiatively inefficient mode
of growth associated with efficient production of radio jets that can
heat gas in a quasi-hydrostatic hot halo. The accretion rate in this
phase is modelled assuming Bondi accretion using the isothermal
cooling flow solution of \citet{nulsen_fabian:00}. We then assume that
the energy that effectively couples to and heats the hot gas is given
by $L_{\rm heat} = \kappa_{\rm heat} \eta \dot{m}_{\rm radio} c^2$,
where $\dot{m}_{\rm radio}$ is the accretion rate onto the BH,
$\eta=0.1$ is the assumed conversion efficiency of rest-mass into
energy, and $\kappa_{\rm heat}$ is a free parameter of order unity.

\section{The Global Evolution of Galaxies and Black Holes}

We normalize the free parameters in our models to reproduce a key set
of observational quantities at $z\sim0$, such as the stellar mass
function, the cold gas fraction in disk galaxies, and the stellar
metallicity (S08). We show our model predictions and compare with
observations for a broad range of low redshift galaxy properties in
S08.

\begin{figure}
\plottwo{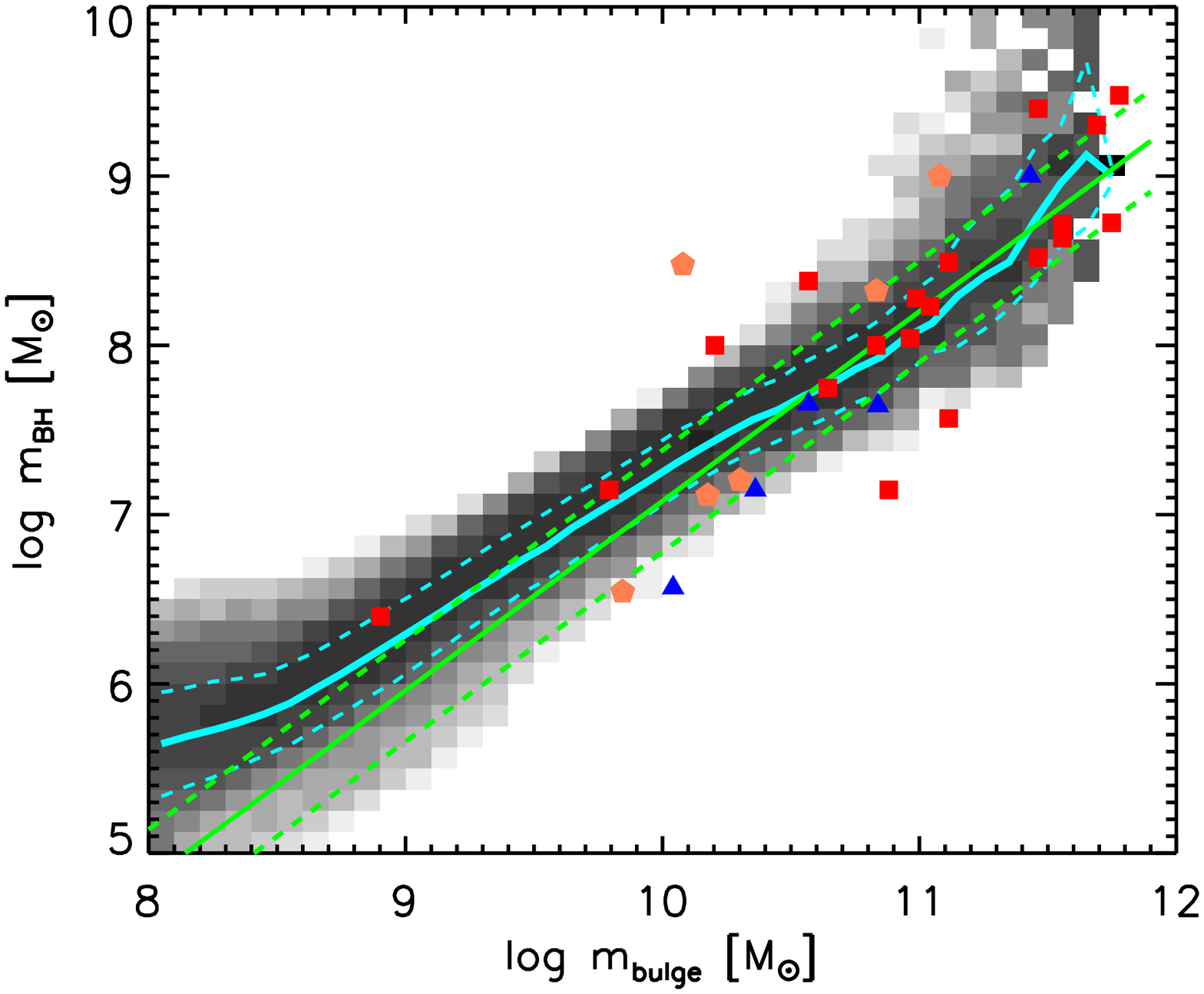}{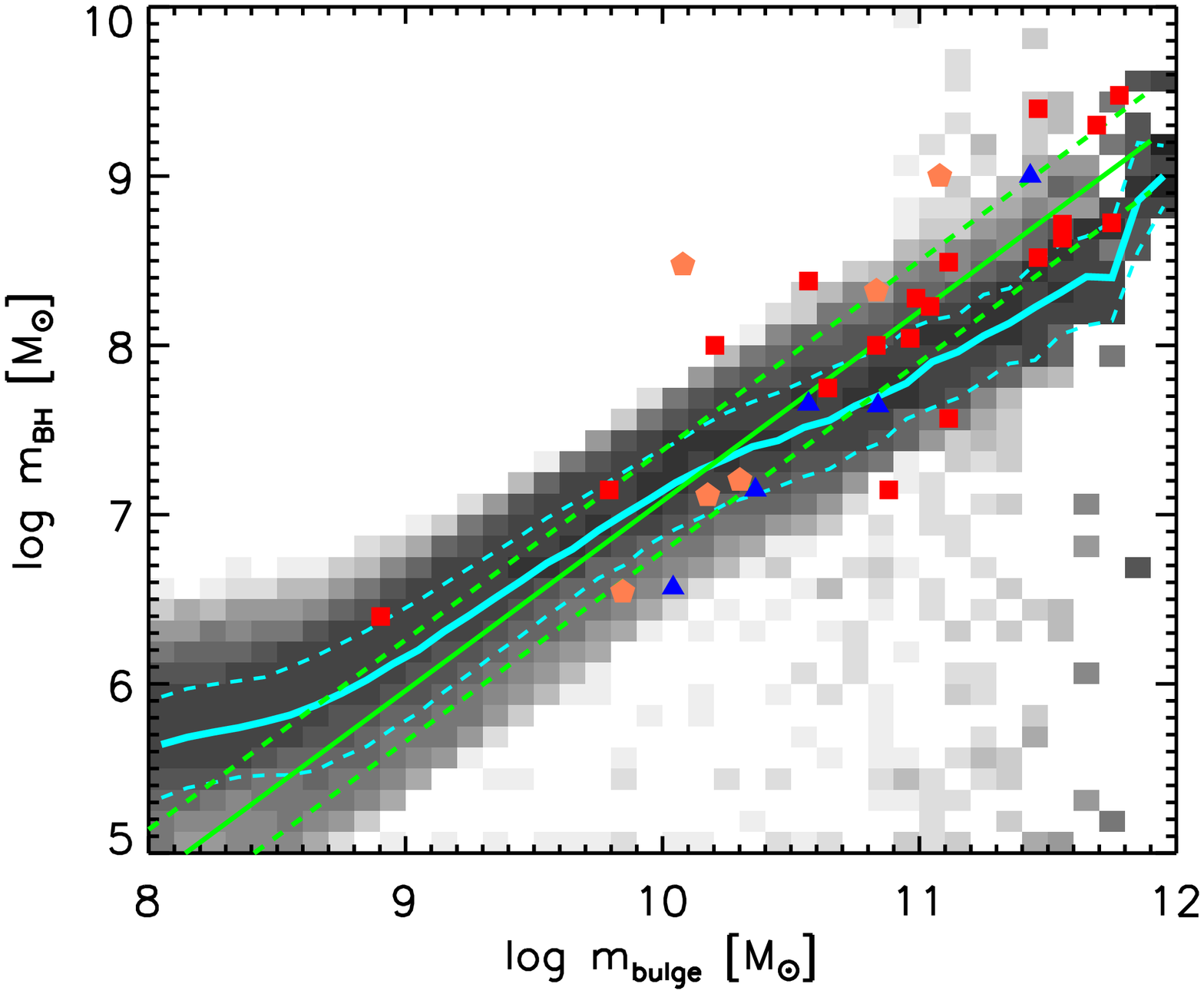}
\caption{The predicted relationship between bulge mass and black hole
  mass (grey shading indicates the conditional probability $P(m_{\rm
    bh}|m_{\rm bulge})$; light blue solid and dashed line shows the
  median and 16th and 84th percentiles) compared with the observed
  relation from \protect\citet[][green lines]{haering:04}. Symbols
  show the measurements for individual galaxies from
  \protect\citet{haering:04}.}
\label{fig:mbh}
\end{figure}

The relationship between galaxy mass and BH mass is clearly a key
result that our model should reproduce. In our model, this
relationship is set by the depth of the potential well of the galaxy
at the time when the BH forms, which in turn is determined by the gas
fraction of the progenitor galaxies of the last merger. More gas-rich
progenitors suffer more dissipation when they merge, and produce more
compact remnants with deeper potential wells. A deeper potential well
requires more energy, and therefore a more massive BH in order to halt
further accretion and growth. We see from Fig.~\ref{fig:mbh} that our
fiducial model reproduces the observed slope and scatter of the
$M_{\rm BH}$--$M_{\rm sph}$ (black hole mass vs. spheroid mass)
relationship. Our model also predicts that the $M_{\rm BH}$--$M_{\rm
  sph}$ relation should evolve with time. Because more gas-rich merger
progenitors produce remnants with larger black holes, and the galaxies
in our models were significantly more gas rich in the past, galaxies
have larger BH for their spheroid mass than they do at the present
day. This leads to a relatively mild amount of evolution, of a factor
of less than two since $z\sim1$ and a factor of about four since
$z\sim 3$ \citep{hopkins_bhfpth:07}.

\begin{figure}
\plottwo{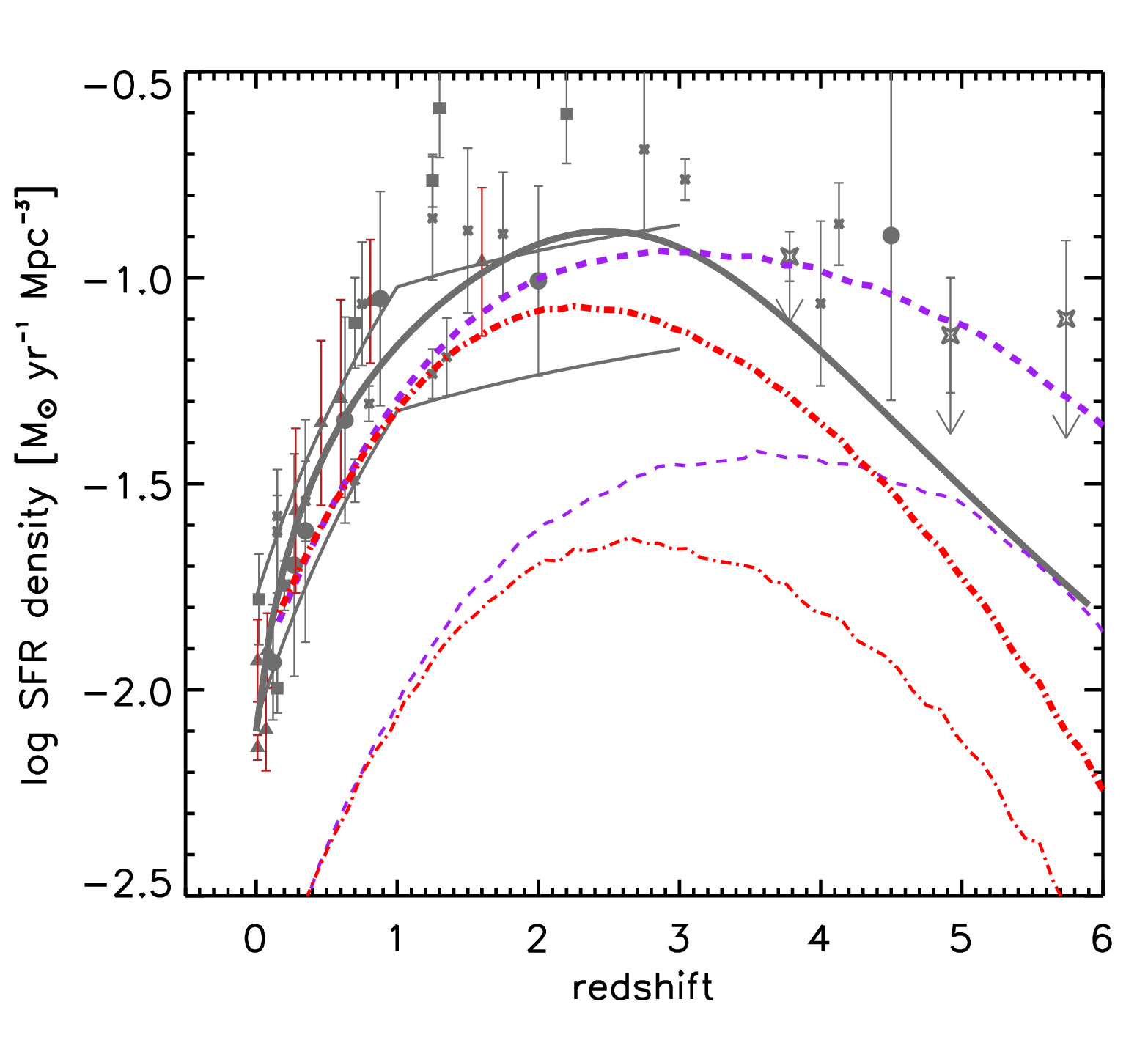}{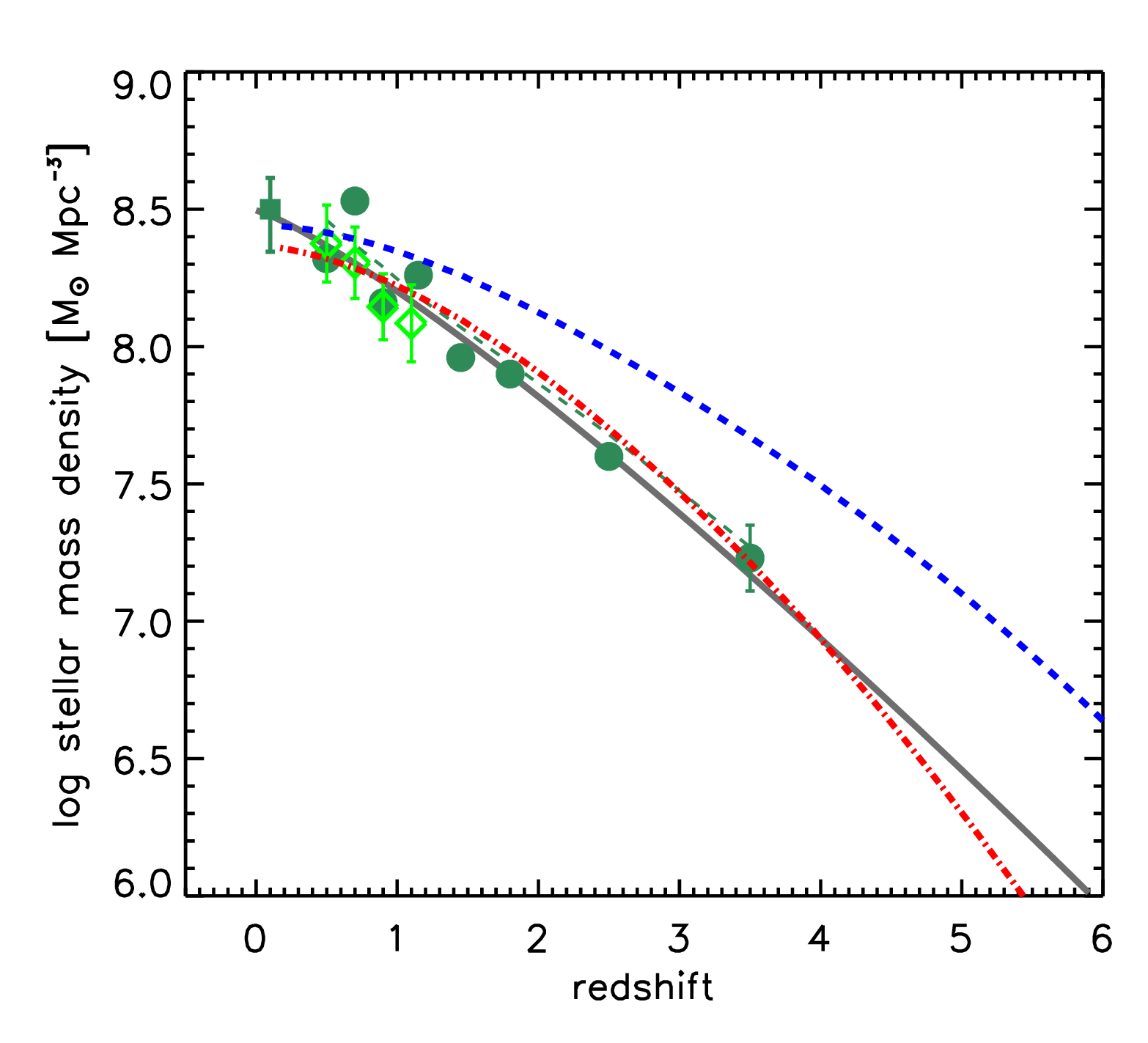}
\caption{Left: Star formation rate density as a function of
  redshift. The upper set of thicker lines shows the total SFR in the
  models, and the lower set of thin lines shows the SFR due to
  bursts. Symbols and solid lines show a compilation of observational
  results, converted to a Chabrier IMF (see S08 for details).  Right:
  The integrated global stellar mass density as a function of
  redshift. Symbols and the solid line show observational estimates
  (see S08).  In both panels, dashed (purple) curves show our
  ``fiducial'' $\Lambda$CDM model, and dot-dashed (red) curves show a
  model with reduced galaxy formation in small mass halos (see
  text). }
\label{fig:global}
\end{figure}

Fig.~\ref{fig:global} shows the global star formation rate density of
all galaxies predicted by our models. We show both the results of our
``fiducial'' model, as well as a model in which galaxy formation has
been suppressed in low-mass halos (we simply do not allow gas to cool
in halos with mass less than $10^{11} M_{\odot}$). We will refer to
the latter model as the ``low'' model for brevity. The results of the
low model are similar to those from models with reduced small scale
power, e.g. with lower $\sigma_8$. We have kept the same values for
the free parameters in both models. Both models fit the data fairly
well from $0<z<1$, and are about 0.15--0.2 dex lower than the
observational compilation of \citet{hopkins_beacom:06} from
$1<z<3$. The fiducial model predicts much more star formation at very
high redshift ($z>3$) than the low model, because much of the star
formation at these redshifts is taking place in low-mass halos.

In the right panel of Fig.~\ref{fig:global} we show the complementary
quantity $\rho_{\rm star}$, the integrated cosmic stellar mass
density. The fiducial model predicts a significantly \emph{earlier}
assembly of stars in galaxies than observations of high redshift
galaxies indicate. However, the low model produces very good agreement
with the stellar mass density as a function of redshift. We note that
this tension in the model results is connected with a possible
inconsistency between the two observational data sets (star formation
rates and stellar masses) that has been noted recently in several
papers
\citep[e.g.][]{hopkins_beacom:06,fardal:07,wilkins:08,dave:08}. One
possible resolution of this tension can be obtained if the stellar IMF
has changed with time, and was more top-heavy at high redshift.

\section{Evolution of ``Bright Mode'' AGN}

\begin{figure}
\plotone{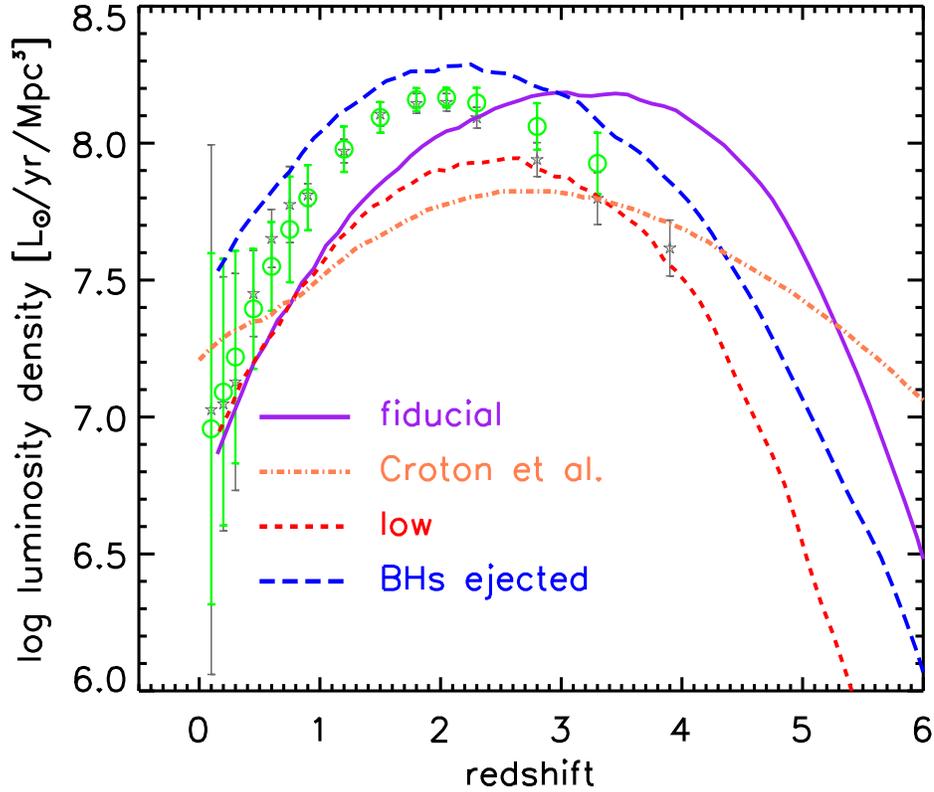}
\caption{The bolometric luminosity density of ``bright mode'' AGN
  (quasars) as a function of redshift. Symbols with error bars show
  the observational estimates from
  \protect\citet{hopkins_qsolf:07}. Lines show model predictions from:
  our fiducial model (solid), our ``low'' model (short-dashed), a
  model in which black holes are ejected after every merger (long
  dashed), and the model of \protect\citet{croton:06}. }
\label{fig:qsolumdens}
\end{figure}

Fig.~\ref{fig:qsolumdens} shows the integrated bolometric luminosity
density of ``bright mode'' BH accretion in our models, compared with
observational estimates from quasar luminosity functions
\citep{hopkins_qsolf:07}. We see that in our fiducial model, QSO activity peaks
at too high a redshift, and falls off too rapidly at low redshift. The
``low'' model predicts a rise and fall in QSO activity that is in
somewhat better agreement with the data, but still does not have quite
the right shape. For comparison, we plot the bright mode accretion
rate in the models of \citep{croton:06}, which are similar to our
fiducial models in many respects, but treat BH growth and AGN feeding
differently. We see that this model also fails to reproduce the
observations: QSO activity peaks too early, and does not fall off
rapidly enough at low redshift.

This result is surprising, because \citet{hopkins_cosmo1:08} found
good agreement with these same observations, using the same model for
BH growth that we have implemented here. We checked that the merger
mass function predicted by our model agrees with that in the Hopkins
et al. models. The gas fractions of our merger progenitors also agree
well. Eventually we realized that an important difference in the
assumptions made by \citet{hopkins_cosmo1:08} and those contained in
the models presented here was that, in the Hopkins et al. model, the
black holes were allowed to grow from the seed mass to the final mass
\emph{in each merger}. This effectively neglected the pre-existing BH
that grew in earlier mergers. Put another way, in our models, by
$z\sim 1$--2, there is already a significant amount of mass in
spheroids in place, each containing a massive black hole. If the
pre-existing BH is already above the critical mass discussed above,
then no AGN activity takes place. Thus, in our models, many late
mergers do not trigger any AGN activity.

As a check, we re-ran our model, assuming that the black holes are
ejected from the galaxy at the end of every merger. The bright mode
accretion history now agrees much better with the observations at
least in shape, although the normalization is a bit high. Although
this model may seem artificial, it is possible that in some cases
black holes are actually ejected by the gravitational rocket
mechanism, which may impart kicks of several hundreds to 1000 km/s to
merging BH \citep[][and references therein]{volonteri:07}. However,
this ejection of massive BH can clearly have a problematic impact on
the predicted $M_{\rm BH}$--$M_{\rm sph}$ relation, as shown in
Fig.~ref{fig:mbh}. However, the model shown here is extreme in that BH
are ejected from every galaxy after every merger. It is possible that
a more physical implementation of the gravitational rocket mechanism
could improve the BH accretion history without leaving behind too many
massive spheroids with no BH, or with very low mass BH, which are not
observed.

Another possible resolution to this problem could perhaps be obtained
if spheroids are not formed as efficiently in high redshift major
mergers as we have assumed here. Because in our models, the properties
of the spheroid regulate BH formation, if spheroids form later, then
BH accretion will occur later as well. There are indications from a
detailed analysis of the hydrodynamic merger simulations that gas-rich
mergers may often produce disk-like, rather than spheroidal, remnants
\citep{robertson:06b,hopkins:08}. Because our disks tend to be more gas
rich at high redshift, this effect would indeed shift spheroid
formation to later times. We are currently working on implementing
this effect in our models.

\section {Summary and Conclusions}

We have presented some first predictions from new models for the joint
evolution of galaxies, black holes, and AGN. Our models differ from
other semi-analytic models in the literature in that we have
implemented self-regulated BH growth based on the results of high
resolution hydrodynamic simulations of galaxy mergers.

We showed that our fiducial model does a fairly good job of
reproducing the global star formation history, but overproduces the
stellar mass in place at high redshift. A ``low'' model with cooling
suppressed in low-mass halos, or a model with reduced small-scale
power, produces good agreement with the stellar mass density history
but does not agree as well with the star formation history. This
reflects an internal tension between these two data sets, which may be
due to systematic errors arising from a time-varying stellar Initial
Mass Function. 

We found that our model, which does fairly well at reproducing galaxy
observations at both low and high redshift, fails to reproduce the
form of the BH accretion history as traced by the bolometric quasar
luminosity density. In our models, QSO activity peaks at too high a
redshift, and falls off too rapidly at low redshift. The models of
\citet{croton:06} also fail to reproduce this behavior. We suggested
two possible resolutions to this problem: 1) Black Holes are (at least
sometimes) ejected from their host galaxy following a merger or 2)
major mergers at high redshift are not as efficient at producing
spheroidal remnants as we have assumed. 

\acknowledgements 

I thank the organizers of the Subaru meeting for putting on an
excellent conference and for inviting me. I thank my collaborators,
P. Hopkins, T. Cox, B. Robertson, Y. Li, and L. Hernquist for allowing
me to present our unpublished work here. I also thank my collaborators
X. Zheng, E. Bell, H.-W. Rix, K. Jahnke, and F. Fontanot for
interesting discussions. I am grateful to D. Croton for providing his
model predictions in electronic form, and for useful discussions.





\end{document}